\documentclass{notes}
\pdfoutput=1
\usepackage[utf8]{inputenc}
\usepackage{tikz}
\usetikzlibrary{shapes,arrows}
\usepackage{graphicx}
\graphicspath{ {} }
\usepackage{hyperref}
\usepackage{listings}
\title{Interstitial Content Detection}
\author{Elizabeth Lucas, Mozilla Research}
\date{August 2017}

\begin{document}

\maketitle

\begin{abstract}
% ASAJ: reworded
Interstitial content is online content which grays out, or otherwise obscures the main page content. In this technical report, we discuss exploratory research into detecting the presence of interstitial content in web pages. We discuss the use of computer vision techniques to detect interstitials, and the potential use of these techniques to provide a labelled dataset for machine learning.
\end{abstract}

\section{Introduction}
% Not sure if I should move the github link to a further reading section? Or where? Seems out of place here. 
The structure and underlying nature of content in the web is fundamentally different than most rigorously structured data, and often requires deviating from the traditional approaches of recognizing patterns in more heavily structured data. Within the types of content on the web, interstitials are of interest due to their interrupting of the user's web experience.

This report represents the preliminary research necessary to explore the structure of interstitial content, and the beginnings of a machine learning application to assist with our understanding of web content and interstitials. The scripts used for data collection and evaluation are available~\cite{InterstitialGithub}.

% ASAJ: An overview and general chat should go here

\subsection{Definitions}
For the purpose of this research project, `interstitials', or `interstitial content', are defined as online content, often advertisements or other promotional content, which grays out or otherwise obscures the main page content. These interstitials often require the user to interact in order to return to the main content, interrupting the user's experience.

\begin{figure}[h!]
    \centering
    \includegraphics[width=.9\textwidth]{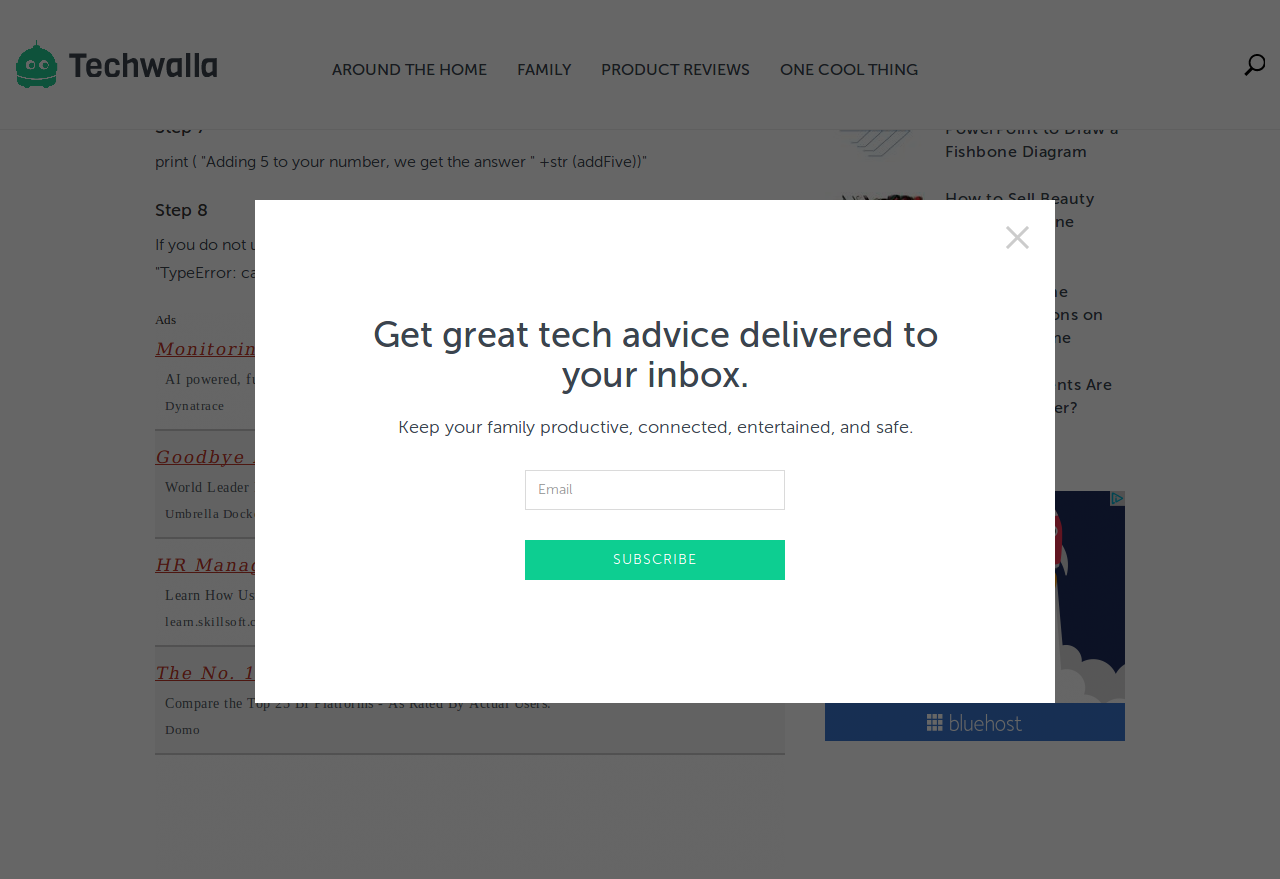} %\hfill
    \caption{An example of an interstitial}
    \label{fig:interstitial}
\end{figure}

`Servo' refers to the Servo browser engine, sponsored by Mozilla Research \cite{Servo}. Written in the Rust programming language, this modern parallel browser engine aims to improve performance, security, modularity, and parallelization. Future work will involve eventually bringing interstitial ad detection into the Servo browser engine itself. 

The Document Object Model, or DOM, is an interface for interacting with webpages at the top level, and allows developers and users to retrieve information about the structure of the page \cite{DOM}. The DOM enables access and manipulation of the page nodes and objects and is the point of contact for web scripts.
%Other definitions or clarifications? 
\subsection{Proposed Method}
\label{proposal}
When approaching a problem from a pattern recognition standpoint, one must have a dataset and a set of features about which to analyze the data. From there, one is able to begin work on the machine learning pieces: deciding on a method, and adjusting parameters as necessary to attain the desired result. Since there is no prior work in this area, the following steps needed to be taken:
\begin{itemize}
    \item Find and label a dataset
    \item Generate features
    \item Generate a model
    \item Create a response (to the interstitials)
    \item Integrate model and response into Servo
\end{itemize}

As of the writing of this report, a baseline for the first three has been established. Future work, as outlined in Section~\ref{sec:future}, will focus on improving the current model through a larger labeled dataset and better, more robust features, as well as writing and integrating the response into Servo.

\subsection*{Acknowledgements} 

Over the course of this research project I have received invaluable support from the Mozilla community including Mozilla Research, the Servo team, and the summer intern cohort. Special thanks to Alan Jeffrey, Jack Moffitt, Manish Goregaokar, and Simran Gujral for continued help and support throughout the summer. 

\section{The Dataset}

The initial list of websites consisted of a small collection of personally curated sites mainly consisting of news sites and tech blogs that the author and colleagues discovered during their usual browsing habits. From that small collection, we were able to create the prototype for a computer vision system as inspired by a related system on recognizing license plates in photos of cars to detect interstitial content \cite{licenseplates}. We leveraged the Common Crawl dataset to obtain lists of valid URLs, as well as potentially using the crawling data in the future for features. Producing an accurate computer vision system will automate the process of labeling the websites from Common Crawl by whether they have an interstitial, thus expanding the amount of labeled data available.

%http://www.iaeng.org/publication/WCECS2012/WCECS2012_pp199-203.pdf

\subsection{Computer Vision}

%ASAJ: passive voice!

The problem of finding and labeling a dataset led to an exploration of computer vision methods to quickly build out a dataset using the visual features of interstitials. Using a small list of hand-labeled sites to test, we used Selenium to create a headless browser that `scrolled' through the sites and took screenshots at regular intervals \cite{Selenium}. From there, we performed hough line transformations through OpenCV on a grayscale version of the images in order to detect straight lines in the image, hopefully representing the bounding boxes of an interstitial.
% screenshots -> hough lines -> filtering -> heuristics 

\subsubsection{Hough Lines}
OpenCV's hough lines implementation was leveraged in order to identify lines in the screenshots, and basic filtering and parameter adjusting was used to improve results \cite{HoughLines}. The images from each part of the scroll through the website were clustered together and passed through a grayscale filter as well as the OpenCV `canny' function, identifying the edges in the images, before being passed through the hough line filter to identify lines from the edges. The $(\rho, \theta)$ tuples used to represent the lines were stored for later examination by the scoring system listed in the next section. 

Filtering the tuples for lines with a $\theta$ of $1.57 \pm 0.005$ for horizontal lines or zero for vertical quickly removed any diagonals caused by artifacts in the images, and using a threshold of 400 helped to remove false positives caused by lines of text or otherwise linearly formatted content in the images. These numbers proved useful for the hand-labeled sites, but further adjustments may be necessary once more ground truth is available. 

\subsubsection{Heuristics}

The amount of filtered lines that remained constant over several images, plus running a basic pixel-level diff over the middle of the image provided the basis for the heuristics. This method was by no means foolproof, often resulting in false positives due to static content placement in websites. 

The majority of the heuristics consisted of finding lines that persisted over multiple images. This helped to remove lines that were found due to noise in the images, as well as navigation or menu bars that would only appear at certain parts of the webpage. If there were were more than one horizontal and vertical line that persisted over more than four images, the website received the maximum amount of points from the hough line heuristics. If there were more than two images in the batch where the middle of the images were the same (hopefully pointing to an interstitial that persisted through scrolling), additional points were awarded. 

\begin{lstlisting}[label={list:first},caption=Heuristics pseudocode]
calculateConfidence(lineCandidates, numdiffs):
    maxConf = 9
    confidence = 0
    results = countFrequency(lineCandidates) 
    if (results):
        pruned = {k:v for k,v in results if v > 1}
        prunedMore = {k:v for k,v in results if v > 4}
        Vert = {(rho,theta):v for (rho,theta),v in pruned if theta  == 0}
        Horz = {(rho,theta):v for (rho,theta),v in pruned if (abs(1.57 - theta) < .005)}
        VertPruned = {k:v for k,v in Vert if v > 4}
        HorzPruned = {k:v for k,v in Horz if v > 4}
        if(pruned):
            confidence += 1
        if(prunedMore): 
            confidence += 2 
        if(len(Vert) > 1 and len(Horz) > 1): 
            confidence += 2
        if(len(VertPruned) > 1 and len(HorzPruned) > 1):
            confidence += 2
        if(numdiffs > 2):
            confidence += 2
    return confidence / maxConf
\end{lstlisting}

\subsubsection{Results}

Overall, counting a `no' as below 0.3 and a `yes' as above 0.75 yielded few incorrectly labeled websites given the small samples of hand-labeled data, but the vast majority of data did not reach either threshold and could not be accurately sorted using this method. These preliminary results were limited due to several problems both with Selenium and the heuristics selected.

One major issue that was discovered with using Selenium was the false positives generated by static webpages. By fetching the scroll height of the page, we dynamically changed how much selenium would scroll each time in order to ensure the script is capturing of most of the page and to stop collecting screenshots should the headless browser reach the end of the page. However, due to some inconsistencies with how this variable is set across webpages, some sites would have several extra screenshots of bottom of the page. This threw off the heuristics, as the hough lines would `persist' in the images but in reality would be from the same part of the webpage. 

Using the basic pixel-level diff for the middle of the image had its own problems, as a non-trivial number of the interstitials featured animated content which would lead to the diffs not recognizing that the content was the same as they were capturing different frames of the animation or video. The diffs also gave points to websites that incorrectly had duplicate images of the same part of the webpage due to the aforementioned problem with retrieving the scroll length in Selenium, lending additional confidence to a false positive. 

 Expanding on these approaches to include more descriptive heuristics may lead to an automated, albeit slow, way of labeling websites. Collecting the screenshots and running them through the heuristics took about an hour per 100 webpages. 

\subsection{Common Crawl}
We selected Common Crawl (CC) as the source for the dataset because of the sheer amount of website data available. Although only the URLs were used out of the available crawl data, there were billions of unique websites listed in this dataset that could be accessed through the CC Index. However, the formatting of the index was not optimal for pulling individual pages without knowing the domain names beforehand. To circumvent this problem, we downloaded the 2015 index, stored as a b-tree, in its entirety and manually navigated through the binary and pulled out the page URLs \cite{CCIndexBlog}. The full URLs were then chopped to only the domain, such as `http://foobar.com' from `http://foobar.com/2017/8/ArticleName.html'. We then used the domain names to query the more recent index from June 2017 through the API provided by CC for more up-to-date page URLs. For convenience the first page of results was pulled from the API and a random URL was selected to be included in the dataset.
%Overview of common crawl. Talk about cleaning the raw 250 gig data dump from 2015. Using the raw data, stripping for domain names then using the index to pull data from the 2017 common crawl index. 

\section{Feature Engineering}
The task of finding discriminating features in an area that has not been previously explored was an exercise in transferring domain knowledge from web experts into data collection. The rest of the Common Crawl data was not utilized due to worries that their crawlers may have received a different version of the pages visited than a headless browser would, and may not contain any advertisements, let alone the same ones as we may find in selenium. Instead, features were initially selected based on the amount and accessibility of data that could be collected through Selenium. 

\subsection{HTML Features}
We quickly discovered problems when attempting to collect complete webpage data, as Selenium does not directly allow for full downloads of the source files. However, we were able to obtain the top level HTML for each website as selenium scrolled through the sites. With this data, we extracted the pairs of HTML elements and attributes as tuples and counted their frequency as a percentage of the total count of element and attribute pairs present in the HTML. 

\subsection{DOM Features}
%Since this is mostly future work....remove? \& restructure feature engineering area? 
Future work will consider the viability and effectiveness of including DOM information into the feature vector. Slight variations in the naming conventions of parts of the web pages may force abstraction or generalization of the data in order to avoid detrimental granularity.

\section{Machine Learning}

The proof-of-concept pipeline was completed by training and testing the validity of a machine learning model on the data labeled by the computer vision system. At this stage, choosing a specific model was of less importance than completing the circuit from raw data to results. 

\subsection{Method}
Scikitlearn was used to train both a default Support Vector Machine (SVM) and to experiment with label spreading due to the limited amount of hand-labeled data available \cite{SKTLearn}. Due to the final timeline of the project as well as the lack of ground truth information available, the models were used as a proof-of-concept to complete the pipeline from the Selenium crawl to a `yes' or `no' decision. The only feature included in this first pass was the list of frequencies of the element and attribute pairs from the HTML source. 
%Scikitlearn. Looked at SVM and label spreading.
\subsection{Results}
Using a small sample of around 200 web pages that had been passed through the prototype of the computer vision system, we were able to train and test a basic SVM. From those 200 pages, roughly half were unable to be labeled, and the remaining pages plus the hand-labeled set were split into the training and testing data, as outlined in table \ref{table:1} below: \hfill
\begin{table}[h!]
\centering
\begin{tabular} { | c | c | c | c | c |}
\hline
& precision & recall & f1-score & support \\
\hline
no & 0.40 & 1.00 & 0.57 & 21 \\
\hline
yes & 0.00 & 0.00 & 0.00 & 32 \\
\hline
avg / total & 0.16 & 0.40 & 0.22 &  53 \\
\hline
\end{tabular}\\
\caption{Confusion Matrix}
\label{table:1}
\end{table}

\section{Future Work}

Further research on this project will focus on improving the first three bullets mentioned in Section \ref{proposal} (dataset, features, and modeling), establishing a response to the identified interstitials, as well as integrating the final model and response into Servo.

\label{sec:future}
\subsection{Dataset Improvement}
The main limiting factor on the dataset is the complete lack of ground truth labeling, which desperately needs to be improved. This can be accomplished either by labeling data manually, or improving the computer vision system to drastically reduce the error rate to within acceptable ranges. 

Further research is needed in determining what data can be collected that will also be present during regular browsing in Servo; the Common Crawl data, after careful comparison to an end user's experience on the same web pages, may be utilized at a later date. This also applies to any DOM information that we may be able to extract through Selenium or Servo in order to collect more overall data on these web pages. 

\subsection{Feature Improvement}
After collecting additional data, we can focus on picking more discriminatory features from the dataset. More research is required to understand the usefulness of any of the data as features for the machine learning models.  

\subsection{ML Improvement}
The current model's performance is limited by both the dataset and the feature space, and further work should focus on finding a model that works well with the feature space. After finding an accurate model that can determine whether a page has an interstitial, the next step will be to use this model as a springboard with which we can begin to identify the individual DOM elements or scripts that constitute the interstitial itself. 

\subsection{Servo Integration}
The final piece of this project is to integrate a model into Servo which can detect the exact DOM elements that are creating the interstitial so that the browser can react to the presence of these interstitials. This model should be lightweight and fast, to avoid any unnecessary slowdown on the browsing engine. 

\clearpage
\bibliographystyle{plain}
\bibliography{main}

\end{document}